\documentclass[aps,pra,twocolumn,superscriptaddress]{revtex4-1}
\usepackage{xcolor,graphicx,ulem}
\usepackage{amsmath,amssymb}
\usepackage[colorlinks=true,urlcolor=blue,citecolor=blue,linkcolor=magenta]{hyperref}

\begin{document}

\title{Entanglement verification of noisy N00N states}

\author{M. Bohmann}\email{martin.bohmann@uni-rostock.de}
\affiliation{Institut f\"ur Physik, Universit\"at Rostock, Albert-Einstein-Str. 23, D-18051 Rostock, Germany}

\author{J. Sperling}
\affiliation{Clarendon Laboratory, University of Oxford, Parks Road, Oxford OX1 3PU, United Kingdom}

\author{W. Vogel}
\affiliation{Institut f\"ur Physik, Universit\"at Rostock, Albert-Einstein-Str. 23, D-18051 Rostock, Germany}

\begin{abstract}
	Entangled quantum states, such as N00N states, are of major importance for quantum technologies due to their quantum-enhanced performance.
	At the same time, their quantum correlations are relatively vulnerable when they are subjected to imperfections.
	Therefore, it is crucial to determine under which circumstances their distinct quantum features can be exploited.
	In this paper, we study the entanglement property of noisy N00N states.
	This class of states is a generalization of N00N states including various attenuation effects, such as mixing, constant or fluctuating losses, and dephasing.
	To verify their entanglement, we pursue two strategies: detection-based entanglement witnesses and entanglement quasiprobabilities.
	Both methods result from our solution of so-called separability eigenvalue equations.
	In particular, the entanglement quasiprobabilities allow for a full entanglement characterization.
	As examples of our general treatment, the cases of N00N states subjected to Gaussian dephasing and fluctuating atmospheric losses are explicitly studied.
	In any correlated fluctuating loss channel, entanglement is found to survive for non-zero transmissivity.
	In addition, an extension of our approach to multipartite systems is given, and the relation to the quantum-optical nonclassicality in phase-space is discussed.
\end{abstract}

\date{\today}

\maketitle

\section{Introduction}

	In quantum information, certain quantum states are of particular interest as they allow for performing tasks beyond classical limitations.
	Among these states is the prominent, two-mode entangled N00N state \cite{Boto2000}.
	Due to its phase sensitivity, this state is of great interest in the fields of quantum metrology \cite{Giovannetti2011}, quantum imaging \cite{Kolobov2007}, and quantum lithography \cite{Boto2000}.
	Even though the experimental preparation of N00N states is rather challenging, several realizations have been reported, e.g., \cite{Mitchell2004,Walther2004,Afek2010,Wang2011,Kruse2015,Lebugle2015}.
	
	In order to benefit from the distinguished characteristics of quantum states when employed in quantum technologies, it is crucial to assess their performance under realistic conditions.
	In other words, one needs to investigate in which scenarios of perturbations one could expect certain quantum enhancements over classical strategies.
	For such an evaluation, one has to take into account various deficiencies, such as imperfect state preparation and manipulation, loss effects, phase noise, and others.
	Therefore, suitable experimental and theoretical tools are required to characterize desirable quantum properties.

	A fundamental quantum property of a compound quantum system is entanglement.
	It has been the central idea in discourses on the principles of quantum mechanics \cite{Schroedinger,EPR}, and it is the basis for various applications in quantum information \cite{Nielsen2000}.
	Although there exist a number of methods for the verification of particular kinds of entanglement \cite{Guehne2009,Horodecki2009}, it is a challenging task to detect entanglement in general.
	Among the vast variety of successfully applied entanglement probes, the method of entanglement witness is one of the most established ones \cite{Horodecki1996,Horodecki2001}.
	Entanglement witnesses are a class of linear operators whose expectation values are nonnegative for separable states but can become negative for entangled states.
	Because of the success of entanglement witnesses, their construction and optimization have been extensively studied, e.g., \cite{Lewenstein2000,Toth2005,Hyllus2006,Sperling2009,Sperling2013,Shahandeh2017}.

	One way to systematically construct entanglement witnesses is based on the separability eigenvalue problem (SEP) \cite{Sperling2009,Sperling2013}.
	Once the SEP is solved for a given linear operator, one is able to construct an optimal witness.
	In the case of multipartite entanglement, this allowed us to experimentally characterize various forms of entanglement beyond earlier limitations \cite{Gerke2015,Gerke2016}.
	Furthermore, this technique has been generalized to quantify entanglement \cite{Sperling2011} and to infer entanglement in systems of indistinguishable particles \cite{Reusch15}.

	Although entanglement witnesses are an effective tool, each witness resembles a sufficient entanglement condition only.
	This means, one cannot conclude the separability of a state by applying a single or few entanglement witnesses.
	A way to overcome this problem is to express the state in terms of so-called entanglement quasiprobabilities \cite{Sperling2009QP}.
	In such a quasiprobability representation, the state is expanded in a separable-state-diagonal form and separability is unambiguously certified by its nonnegativity.
	In contrast to the witnessing approach, any occurring negativity in the entanglement quasiprobabilities provides a necessary and sufficient entanglement verification.
	For example, entanglement quasiprobabilities have been applied to two-mode squeezed-vacuum states \cite{Sperling2012}.

	In this paper, we study the entanglement property of noisy N00N states.
	This generalization of N00N states takes into account various noise effects, such as mixtures, fluctuating losses, and dephasing.
	The SEP is solved for this entire family of states and its relation to witnesses and entanglement quasiprobabilities is shown.
	For the case of Gaussian dephasing and atmospheric losses, we explicitly provide and discuss entanglement witnesses and entanglement quasiprobabilities.
	In addition, we also outline a generalization to witness multimode entanglement, and we compare our entanglement quasiprobabilities to another fundamental quasiprobability representation in quantum optics.

	The paper is organized as follows.
	In Sec. \ref{ch:State}, we define the class of noisy N00N states.
	The SEP and its relation to entanglement witnesses and entanglement quasiprobabilities is considered in Sec. \ref{sec:SEP-methods}.
	In Sec. \ref{ch:SimpleSolution}, we solve the SEP problem for noisy N00N states.
	Explicit examples of the entanglement verification are given in Sec. \ref{ch:application}.
	In Sec. \ref{ch:Generalization}, we discuss multipartite generalizations and comparisons with phase-space quasiprobabilities.
	We summarize and conclude in Sec. \ref{ch:Summary}.

\section{Noisy N00N states}\label{ch:State}

	In this section, we define the class of noisy N00N states, introduced in Ref. \cite{Bohmann2015}.
	As noisy N00N states take into account various imperfections, they provide a more realistic representation of such states from the experimental point of view.
	Specifically, we focus on the influences of dephasing and fluctuating losses, as those effects are of particular importance in the continuation of this work.

	The ideal N00N state with a photon number $N$ reads as
	\begin{equation}\label{eq:N00N}
		|\psi_N\rangle=\frac{1}{\sqrt{2}}\left(|N,0\rangle+|0,N\rangle\right).
	\end{equation}
	In particular, the density operator $|\psi_N\rangle\langle\psi_N|$ includes diagonal contributions of the form $|N,0\rangle\langle N,0|$ and $|0,N\rangle\langle 0,N|$ as well as the interference parts $|N,0\rangle\langle 0,N|$ and its Hermitian conjugate.
	Thus, one can introduce the following generalization \cite{Bohmann2015}:
	Any quantum state $\hat\rho$ which is of the form
	\begin{align}\label{eq:NoisyN00N}
	\begin{aligned}
		\hat{\rho}=&\rho_{00,00} |0,0\rangle \langle 0,0|\\
		&+\sum\limits_{i=1}^{\infty} \Big[\rho_{i0,i0} |i,0\rangle \langle i,0|+\rho_{0i,0i} |0,i\rangle \langle 0,i|\Big]\\
		&+\sum\limits_{i=1}^{\infty}\Big[\rho_{i0,0i} |i,0\rangle \langle 0,i|+\rho_{0i,i0} |0,i\rangle \langle i,0|\Big],
	\end{aligned}
	\end{align}
	is referred to as a noisy N00N state.
	Obviously, a N00N state is in this class, but separable states can also be found in this family, e.g., $|0,N\rangle\langle 0,N|$.
	Moreover, this family of states can be obtained from ideal N00N states \eqref{eq:N00N} by including the imperfections of mixing N00N state with different photon numbers as well as deterministic and fluctuating losses and phase manipulations \cite{Bohmann2015}.

	In this paper we focus on the noise effects of dephasing and fluctuating losses on the entanglement.
	Note that these two effects are also crucial to the quantum-enhanced phase sensitivity \cite{Escher2011,Demkowicz2012} of N00N states.
	In the following, these two noise contributions are explained in more detail.

\subsection{Dephasing}\label{ch:Dephasing}

	The phase of a single-mode state can be deterministically manipulated via the unitary transformation $e^{i\varphi\hat n}$, where $\hat n$ is the photon-number operator.
	Consequently, fluctuating phases describe a two-mode dephasing channel,
	\begin{align}
		\hat\rho\mapsto\Lambda_{\rm deph}(\hat\rho)=&\int_{0}^{2\pi} d\varphi_a\int_{0}^{2\pi} d\varphi_b\, p(\varphi_a,\varphi_b)
		\\&\times\nonumber
		[e^{i\varphi_a\hat n}\otimes e^{i\varphi_b\hat n}]
		\hat\rho
		[e^{-i\varphi_a\hat n}\otimes e^{-i\varphi_b\hat n}],
	\end{align}
	for a phase distribution $p(\varphi_a,\varphi_b)$.
	From an ideal N00N state, we get the noisy N00N state
	\begin{align}\label{eq:dephasing}
	\begin{aligned}
		&\Lambda_{\rm deph}(|\psi_N\rangle\langle\psi_N|)\\
		=&\frac{1}{2}\left[|0,N\rangle\langle0,N|+|N,0\rangle\langle N,0|\right.\\
		&\phantom{\frac{1}{2}}\left.+\lambda|N,0\rangle\langle 0,N|+\lambda^\ast|0,N \rangle\langle N,0|\right],
	\end{aligned}
	\end{align}
	with $\lambda=\int_{0}^{2\pi} d\varphi_a\int_{0}^{2\pi} d\varphi_b\, p(\varphi_a,\varphi_b) e^{i(\varphi_a-\varphi_b) N}$.
	In general, the dephasing yields a reduction of coherence terms, $|\lambda|<1$.
	Moreover, a full dephasing, $p(\varphi)=1/(2\pi)$, yields $\lambda=0$, which causes a complete loss of phase sensitivity and entanglement.
	Note that, however, in the case of full dephasing or no phase information more general quantum correlations might be preserved \cite{Agudelo2013,Sperling2015} and such correlations can still be used for quantum information tasks \cite{ShahandehArXiv}.

	Specifically, we consider the case of Gaussian dephasing in the second mode, which was also studied in Refs. \cite{Sperling2012,Bohmann2015}.
	In this case, $p(\varphi_a,\varphi_b)=\delta_A(\varphi_a)p_B(\varphi_b)$ [$\delta_A$ denotes the Dirac-delta distribution], the phase distribution $p_B(\varphi_b)$ might be modeled by a wrapped Gaussian phase distribution,
	\begin{align}\label{eq:pGauss}
		p_B(\varphi_b)=\sum\limits_{k\in \mathbb Z}\frac{1}{\sqrt{2\pi\delta^2}}\exp\left[-\frac{(\varphi_b+2k\pi)^2}{2\delta^2}\right],
	\end{align}
	where $\delta$ is the width of the initial Gaussian distribution.
	Now, we get the coherence term $\lambda$ in Eq. \eqref{eq:dephasing} as
	\begin{align}\label{eq:lambda}
		\lambda=e^{-\delta^2N^2/2},
	\end{align}
	which decreases for increasing $\delta$.

\subsection{Fluctuating, atmospheric losses}

	As a second scenario of imperfections, we consider losses.
	Specifically, fluctuating losses are the main challenge in free-space quantum communication \cite{Fedrizzi2009,Capraro2012}.
	Importantly, the impact of such atmospheric losses can differ drastically from deterministic loss channels \cite{BSSV2016,BSSV2017}.

	In a first step, we consider the influence of deterministic loss, which gives
	\begin{align}\label{eq:lossy}
		|\psi_N\rangle\langle\psi_N|\mapsto&
		\frac{(T_a T_b)^N}{2}\Big[|N,0\rangle\langle 0,N|+|0,N\rangle \langle N,0|\Big]
		\\\nonumber
		&+\frac{1}{2}\sum\limits_{k=0}^N\binom{N}{k}T_a^{2k}(1{-}T_a^2)^{N-k}|k,0\rangle\langle k,0|
		\\\nonumber
		&+\frac{1}{2}\sum\limits_{k=0}^N\binom{N}{k}T_b^{2k}(1{-}T_b^2)^{N-k}|0,k\rangle\langle 0,k|,
	\end{align}
	with $T_{a},T_{b}\in [0,1]$ being the real-valued field transmission coefficients of the two modes.
	In atmospheric channels, the transmission coefficients do not have a fixed value.
	Rather, they vary in a random manner.
	A quantum optical consistent framework of such fluctuating loss channels has been introduced in Ref. \cite{Semenov2009}.
	In this way, each atmospheric channel is characterized by a probability distribution of the transmission coefficient $\mathcal{P}$, resulting in the moments of the transmission coefficients:
	\begin{align}\label{eq:Tmoments}
		\langle T_a^{m}T_b^{n}\rangle=\int_0^1 dT_a\int_0^1 dT_b\,\mathcal P(T_a,T_b) T_a^{m}T_b^{n}.
	\end{align}
	Different models for $\mathcal P$, representing different atmospheric conditions, have been formulated \cite{VSV2012,VSV2016}.
	
	In the second step, we apply the fluctuating loss to the N00N state.
	This gives a state of the form in Eq. \eqref{eq:lossy}, where the transmission coefficients are replaced with the moments in Eq. \eqref{eq:Tmoments}.
	Namely, we get for a turbulent, atmospheric channel $\Lambda_{\rm atm}$,
	\begin{align}\label{eq:PDTCchannel}
	\begin{aligned}
		&\Lambda_{\rm atm}(|\psi_N\rangle\langle\psi_N|)
		\\
		=&
		\frac{\langle T_a^N T_b^N\rangle}{2}\Big[|N,0\rangle\langle 0,N|+|0,N\rangle \langle N,0|\Big]
		\\
		&+\frac{1}{2}\sum\limits_{k=0}^N\binom{N}{k}\langle T_a^{2k}(1{-}T_a^2)^{N-k}\rangle|k,0\rangle\langle k,0|
		\\
		&+\frac{1}{2}\sum\limits_{k=0}^N\binom{N}{k}\langle T_b^{2k}(1{-}T_b^2)^{N-k}\rangle|0,k\rangle\langle 0,k|.
	\end{aligned}
	\end{align}

\section{Entanglement and the separability eigenvalue problem}\label{sec:SEP-methods}

	In this section, we recall the theoretical framework of the used entanglement detection methods.
	For this purpose, the SEP is considered in a first step.
	After that, we show how its solutions can be used to construct entanglement witnesses and entanglement quasiprobabilities.

\subsection{Separability eigenvalue problem}\label{ch:SEP}

	Let us start with the definition of the SEP for a general Hermitian operator $\hat L$ \cite{Sperling2009,Sperling2013}, which are used for our analysis.
	In the bipartite case, the separability eigenvalue equations are defined as:
	\begin{equation}
		\begin{split}\label{eq:SEP}
			\hat{L}_b |a\rangle =g|a\rangle
			\quad
			\text{and}
			\quad
			\hat{L}_a|b\rangle =g|b\rangle\,
		\end{split}
	\end{equation}
	with the local projections,
	\begin{equation*}
		\hat{L}_a=\mathrm{tr}_A[\hat{L}(|a\rangle \langle a|\otimes\hat{1})] \quad\text{and}\quad \hat{L}_b=\mathrm{tr}_B[\hat{L}(\hat{1}\otimes |b\rangle \langle b|)].
	\end{equation*}
	The values $g$ and the normalized states $|a,b\rangle$ are referred to as the separability eigenvalues and eigenvectors, respectively.
	The SEP, given by the Eqs. \eqref{eq:SEP}, is a coupled eigenvalue problem.
	Thus, solving the SEP is, in general, a nontrivial problem.
	All solutions of the SEP can be collected in a set of pairs,
	\begin{equation}\label{eq:solutionSEP}
		\{(g_i,|a_i,b_i\rangle)\}_i.
	\end{equation}
	So far, we formulated a rather abstract algebraic problem.
	The strength and usefulness of the SEP as well as its solutions for N00N states is presented in the following.

\subsection{Entanglement witnesses}\label{ch:EW}

	We now show how one can use the SEP in order to construct entanglement witnesses \cite{Sperling2009}.
	An entanglement witness is an operator $\hat W$ which yields non-negative expectation values for separable states.
	Thus, $\langle\hat W\rangle<0$ certifies entanglement.
	Any entanglement witness can be written in the form $\hat W{=}w\hat 1-\hat L$, where $w$ is the maximally attainable expectation value for separable states \cite{Toth2005,Sperling2009}.
	Thus, by using the solution of the SEP in Eq. \eqref{eq:solutionSEP}, we can identify $w$ and show that any entanglement witness has the form \cite{Sperling2009}
	\begin{align}\label{eq:W}
		\hat W=\sup_{i}\{g_i\}\hat 1-\hat L.
	\end{align}

	Hence, the solution of the SEP can be used to design witness operators, suited for a desired class of states, to assess the entanglement.
	Also note that this method has been generalized to multipartite entanglement \cite{Sperling2013} and applied to construct optimal witnesses for experimental applications \cite{Gutierrez2014,Gerke2015,Gerke2016}.
	To get customized witnesses for the class of noisy N00N states \eqref{eq:NoisyN00N}, we consider operators $\hat L$ in Eq. \eqref{eq:W} of the form
	\begin{align}\label{eq:testop}
	\begin{aligned}
		\hat{ L}=&L^{(0)} |0,0\rangle \langle 0,0|\\
		&+\sum\limits_{i=1}^{\infty} \Big[ L_i^{(A)} |i,0\rangle \langle i,0|+L_i^{(B)} |0,i\rangle \langle 0,i|\Big]\\
		&+\sum\limits_{i=1}^{\infty}\Big[\gamma_i |i,0\rangle \langle 0,i|+\gamma_i^\ast |0,i\rangle \langle i,0|\Big].
	\end{aligned}
	\end{align}

\subsection{Entanglement quasiprobabilities}\label{ch:QP}

	We have just shown how one can use the solutions of the SEP to construct entanglement witnesses.
	However, an unsuccessful test, $\langle\hat W\rangle\geq 0$, does not allow for concluding separability.
	For this reason, we introduced the method of entanglement quasiprobabilities \cite{Sperling2009QP}, which is also based on the SEP.
	Using this representation, separability is certified if and only if this quasiprobability distribution is nonnegative.
	Conversely, entanglement is uniquely identified by negativities within the entanglement quasiprobability distribution.

	It has been shown in Ref. \cite{Sperling2009QP} that the solutions of the SEP in Eq. \eqref{eq:solutionSEP}---while replacing the test operator with the state, $\hat L=\hat\rho$---allows for expanding the quantum state in terms of separability eigenvectors
	\begin{align}\label{eq:Pent}
		\hat \rho=\sum_k P_{\mathrm{Ent}}(a_k,b_k)|a_k,b_k\rangle\langle a_k,b_k|,
	\end{align}
	where $P_{\mathrm{Ent}}$ is the entanglement quasiprobability distribution.
	Furthermore, a relation between $P_{\mathrm{Ent}}$ and the separability eigenvalues has been formulated.
	Namely, based on the solutions of the SEP of $\hat \rho$, we can construct the following Gram-Schmidt matrix and vector of separability eigenvalues:
	\begin{align}\label{eq:MatrixVector}
		\mathbf{G}=(\langle a_k,b_k|a_l,b_l\rangle)_{k,l}
		\quad
		\text{and}
		\quad
		\vec{g}=(g_k)_k.
	\end{align}
	This allows us to formulate the linear equation
	\begin{align}\label{eq:Gpg}
		\mathbf{G}\vec{p}=\vec{g},
	\end{align}
	where $\vec p$ represents an entanglement quasiprobability distribution.
	By inverting this matrix equation, we obtain the entanglement quasiprobability $\vec{p}$.
	If the matrix $\boldsymbol G$ has the eigenvalue zero, $\vec{p}$ needs to be further optimized over the set of orthonormal eigenvectors $\vec{p}_{0,k}$ to the eigenvalue $0$ of $\boldsymbol G$.
	This gives \cite{Sperling2009QP}
	\begin{align}
		 P_{\mathrm{Ent}}\cong[P_{\mathrm{Ent}}(a_l,b_l)]_l=\vec{p}+\sum_k c_k\vec{p}_{0,k},
	\end{align}
	where $\vec p=\boldsymbol G^{+}\vec g$ ($\boldsymbol G^{+}$ denotes the Moore-Penrose pseudoinverse) and the real coefficients are
	\begin{align}
		c_k=-\frac{\vec{p}_{0,k}^{\rm T}\vec{p}}{\vec{p}_{0,k}^{\rm T}\vec{p}_{0,k}}.
	\end{align}
	
	Therefore, based on the solutions of the SEP of the quantum state $\hat\rho$, we can get the full knowledge on the entanglement properties from the entanglement quasiprobability.
	This is a great asset compared to entanglement witnesses, which only provide a certain test condition rather than a full state characterization.

\section{Solution for noisy N00N states}\label{ch:SimpleSolution}

	We solve the SEP for the whole family of operators of the form of noisy N00N states in Eq. \eqref{eq:NoisyN00N}, likewise the operators in Eq. \eqref{eq:testop}, which can be found in the Appendix.
	This allows us to apply the methods explained in Sec. \ref{sec:SEP-methods}.
	Here, let us focus on a special case to exemplify the general treatment and to avoid a lot of technical details due to the complexity of the full problem.
	We choose to consider the following test operator:
	\begin{align}\label{eq:A}
		\hat L=|N,0\rangle\langle 0,N|+|0,N\rangle\langle N,0|.
	\end{align}
	Note that this operator addresses the interference terms of the noisy N00N states in Eq. \eqref{eq:NoisyN00N}.

	For an efficient solution of the separability eigenvalue equations \eqref{eq:SEP}, we can decompose the separability eigenvector of the first mode as
	\begin{align}
		|a\rangle=&r|0\rangle+\sqrt{1-|r|^2}|N\rangle
	\end{align}
	with a complex $r$ and $|r|\leq 1$.
	The reduced operator---with respect to $|a\rangle$---is
	\begin{align}
		\hat L_a=\sqrt{1-|r|^2}\Big[r|0\rangle\langle N|+r^\ast|N\rangle\langle 0|\Big].
	\end{align}
	Hence, for $0<|r|<1$, we get the eigenvectors to the second equation of the SEP \eqref{eq:SEP} as
	\begin{align}
		|b\rangle=\frac{1}{\sqrt2}\left[|0\rangle\pm\frac{r^\ast}{|r|}|N\rangle\right].
	\end{align}
	Now, it follows that the reduced operator---with respect to $|b\rangle$---reads
	\begin{align}
		\hat L_b=\pm\frac{1}{2|r|}\Big[r^\ast|N\rangle\langle 0|+r|0\rangle\langle N|\Big].
	\end{align}
	The initial vector $|a\rangle$ is an eigenvector of this operator, if and only if $|r|=1/\sqrt 2$.
	In addition, this yields the corresponding separability eigenvalues
	\begin{align}
		g=\pm\frac{1}{2}.
	\end{align}
	Also note that for the trivial choices $|r|=0,1$, we get $g=0$ for the separability eigenvectors $|a,b\rangle$ having the form $|0,0\rangle$, $|0,N\rangle$, $|N,0\rangle$, or $|N,N\rangle$.

	Hence, we get for the here-considered operator $\hat L$ that the maximal expectation value for separable states is $1/2$.
	This allows us to construct a witness of the form given in Eq. \eqref{eq:W}.
	Equivalently, one can state the following: If the state under study $\hat\rho$ is separable, it holds
	\begin{align*}
		 2{\rm Re}(\rho_{N0,0N})=\Big[\langle N,0|\hat\rho|0,N\rangle+\langle 0,N|\hat\rho|N,0\rangle\Big]\leq \frac{1}{2}.
	\end{align*}
	To further generalize this expression, we can use the fact that the separability eigenvalues do not change under local unitary transformations, such as phase transformations.
	Consequently, the above separability constraint generalizes to
	\begin{align}\label{eq:EC}
		|\rho_{N0,0N}|\leq\frac{1}{4}.
	\end{align}
	This means that whenever the absolute value of any interference term in the density matrix exceeds $1/4$, entanglement is verified.

	The above consideration outlines how the SEP can be solved and how its solutions can be used to construct witnesses.
	We also study entanglement quasiprobabilities in the next section.
	We stress that the fully analytical solutions for noisy N00N states \eqref{eq:NoisyN00N} and test operators \eqref{eq:testop} are provided in the Appendix.
	Although our approach to construct measurable entanglement witnesses or necessary and sufficient entanglement quasiprobabilities seem to be more complex at first sight, it is worth emphasizing that the solutions of the SEP are provided here and, therefore, allow for the direct application of our techniques in experiments.

\section{Application to noisy N00N states}\label{ch:application}

	In this section, we study the entanglement properties of the noisy N00N states in Eq. \eqref{eq:NoisyN00N}, based on entanglement witnesses and quasiprobabilities.
	We start our analysis for the pure N00N state.
	Afterwards we proceed to more general noisy N00N states.
	In particular, we study the influence of dephasing and fluctuating losses.
	
\subsection{Pure N00N states}\label{ch:pureN00N}

	For a pure N00N state \eqref{eq:N00N}, we apply the witness condition in Eq. \eqref{eq:EC}.
	Form Eq. \eqref{eq:N00N}, we directly obtain $|\rho_{N0,0N}|=1/2$ and, therefore,
	\begin{align}
		|\rho_{N0,0N}|=\frac{1}{2}>\frac{1}{4}.
	\end{align}
	Hence, we make the well-known observation that the N00N state is entangled, with our simple criterion.

	As an alternative method, we now also consider the entanglement quasiprobabilities, which has not been done before.
	From the solutions in the Appendix for $\hat\rho=|\psi_N\rangle\langle\psi_N|$, we obtain the following pairs of separability eigenvalues and separability eigenvectors:
	\begin{subequations}
	\begin{align}\label{eq:trivial}
		\{
			(0,|0,0\rangle),
			(0,|N,N\rangle),
			(\frac{1}{2},|0,N\rangle),
			(\frac{1}{2},|N,0\rangle)
		\}
	\end{align}
	as well as
	\begin{align}
		\nonumber\{
			&(\frac{1}{2},|\mathfrak{s}_0,\mathfrak{s}_0\rangle),
			(\frac{1}{2},|\mathfrak{s}_1,\mathfrak{s}_1\rangle),
			(\frac{1}{2},|\mathfrak{s}_2,\mathfrak{s}_2\rangle),
			(\frac{1}{2},|\mathfrak{s}_3,\mathfrak{s}_3\rangle),
		\\
			&(0,|\mathfrak{s}_0,\mathfrak{s}_2\rangle),
			(0,|\mathfrak{s}_1,\mathfrak{s}_3\rangle),
			(0,|\mathfrak{s}_2,\mathfrak{s}_0\rangle),
			(0,|\mathfrak{s}_3,\mathfrak{s}_1\rangle)
		\},\label{eq:nontrivial}
	\end{align}
	\end{subequations}
	where $|\mathfrak{s}_n\rangle=(|0\rangle+i^n|N\rangle)/\sqrt{2}$.
	Based on the above solutions, we define the Gram-Schmidt matrix $\boldsymbol G$ and the vector of separability eigenvalues $\vec g$ [cf. Eq. \eqref{eq:MatrixVector}].
	By inversion of Eq. \eqref{eq:Gpg} described in Sec. \ref{ch:QP}, we eventually obtain the entanglement quasiprobabilities for the pure N00N state,
	\begin{align}\label{eq:QuasiProbPure}
		P_{\rm Ent}\cong\frac{1}{4}(0,0,2,2,1,1,1,1,-1,-1,-1,-1).
	\end{align}

\begin{figure}[t]
	\includegraphics[width=\columnwidth]{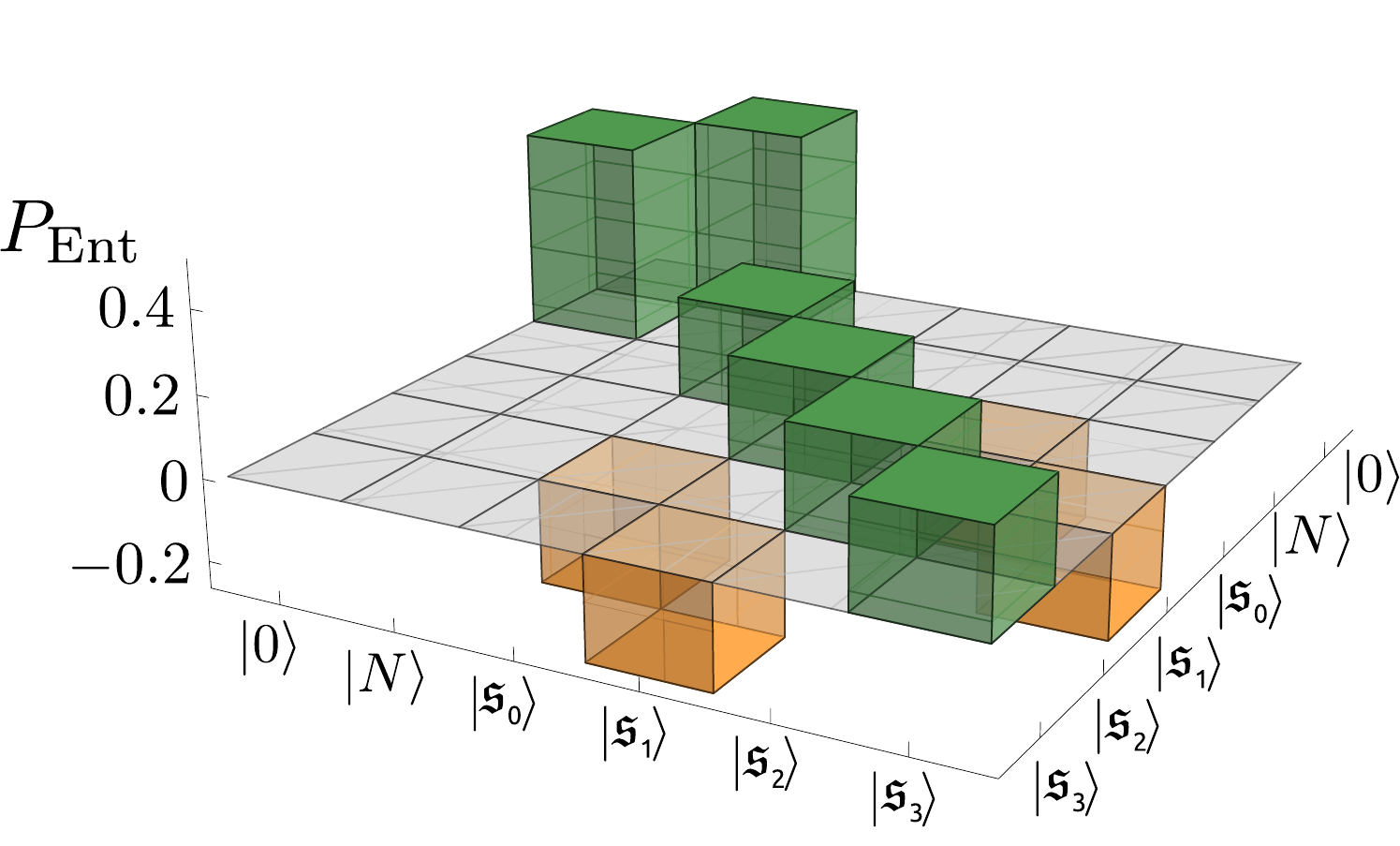}
	\caption{(Color online)
		Entanglement quasiprobability distribution of pure N00N states.
		A patch in the horizontal plane corresponds to a product $|a,b\rangle$ for the given vectors at the axes.
		Note, the solutions are symmetric with respect to subsystems in the considered case.
		The height of the corresponding bar is the value of $P_{\rm Ent}(a,b)$ in Eq. \eqref{eq:QuasiProbPure}.
		The negativities directly reveal the entanglement of the N00N state.
	}\label{fig:pureN00N}
\end{figure}

	The entanglement quasiprobability distribution of the pure N00N state is shown in Fig. \ref{fig:pureN00N}.
	The solutions [Eqs. \eqref{eq:trivial} and \eqref{eq:nontrivial}] of the SEP for a different photon number, $N\neq N'$, is basically identical.
	The only difference is the exchange between $|N\rangle$ and $|N'\rangle$.
	Thus, the entanglement quasiprobability distribution in Eq. \eqref{eq:QuasiProbPure} (Fig. \ref{fig:pureN00N}) is the same for any pure N00N state $|\psi_N\rangle$.
	This quasiprobability distribution in Fig. \ref{fig:pureN00N} shows negativities which doubtlessly display the entangled character of the N00N state.
	It also gives an additional insight into the inseparable character of the distribution $P_\mathrm{Ent}$ over separable states.
	Namely, the negative contributions, $P_\mathrm{Ent}(a,b)<0$, are assigned to the last four states of the solution in Eq. \eqref{eq:nontrivial}, whereas all remaining parts---including those in Eq. \eqref{eq:trivial}---have a positive (i.e., classical) weight, $P_\mathrm{Ent}(a,b)\geq0$.
	
\subsection{Dephasing}

	After characterizing the entanglement of the ideal case, let us study the dephased N00N state in Eq. \eqref{eq:dephasing}.
	In a first step, we apply the witness test based on the violation of the separability condition~\eqref{eq:EC}, which yields the entanglement condition 
	\begin{align}\label{eq:WitnessLambda}
		\lambda>\frac{1}{2}.
	\end{align}
	Therein, the influence of dephasing is characterized by the parameter $\lambda$ [see Eq. \eqref{eq:dephasing}].
	
\begin{figure}[b]
	\includegraphics[width=.8\columnwidth]{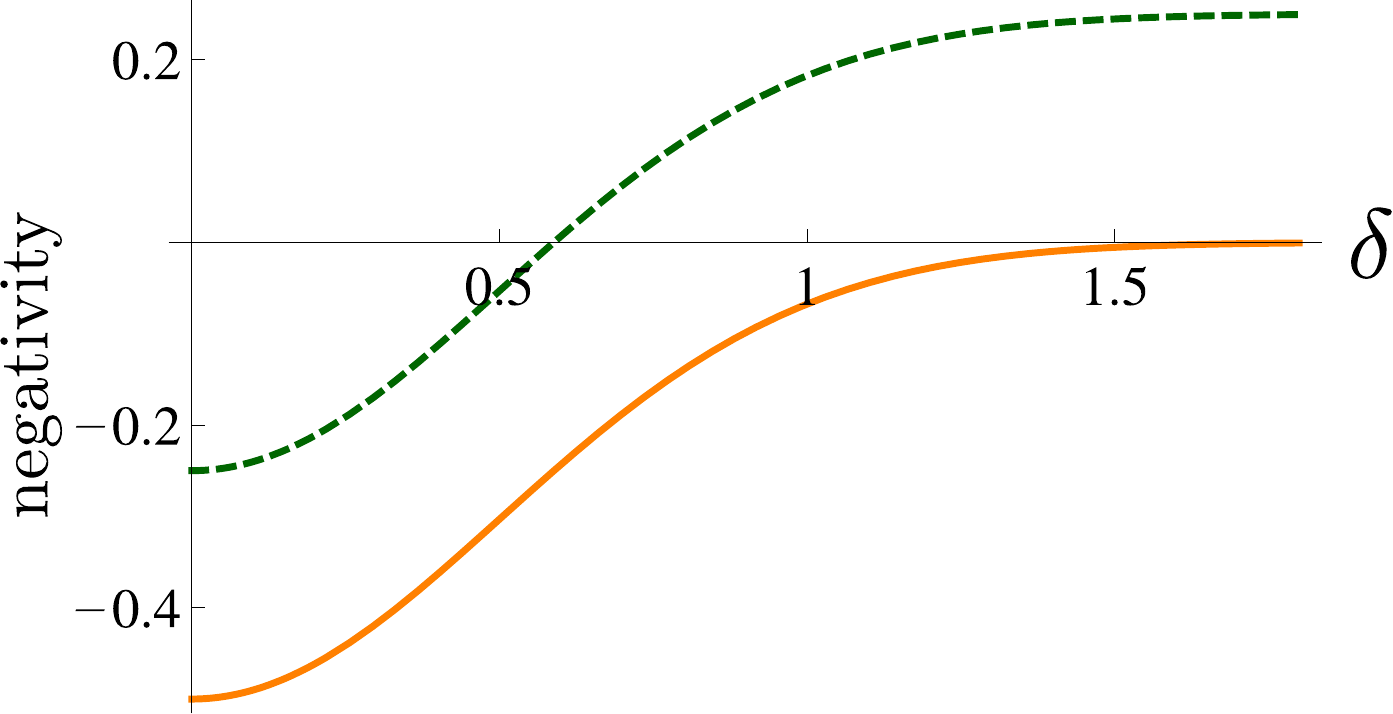}
	\caption{(Color online)
		Entanglement conditions are shown as functions of the strength of Gaussian dephasing $\delta$ for a dephased N00N state with $N{=}2$.
		Negative values indicate the entanglement detection according to the witness condition Eq. \eqref{eq:WitnessDelta} (dashed) and the negativity of the entanglement quasiprobabilities in Eq. \eqref{eq:EQDelta} (solid).
	}\label{fig:depha}
\end{figure}

	We may further consider the specific case of Gaussian dephasing, given by the distribution in Eq. \eqref{eq:pGauss}.
	The strength of the Gaussian dephasing is determined by the width parameter $\delta$, which reduces the interference terms of the N00N state by a factor $\lambda=\exp(-\delta^2N^2/2)$ [cf. Eqs. \eqref{eq:dephasing} and \eqref{eq:lambda}].
	Inequality \eqref{eq:WitnessLambda} verifies entanglement for width parameters with
	\begin{align}\label{eq:WitnessDelta}
		\delta<\frac{\sqrt{2\ln 2}}{N}.
	\end{align}
	Note that the bound scales as the inverse of the photon number, $N^{-1}$.
	This condition is shown by the dashed line in Fig. \ref{fig:depha}.
	Entanglement is certified for low dephasing strengths, whereas no entanglement can be inferred in the case of strong dephasing because N00N states with higher photon numbers $N$ are more sensitive to dephasing.
	Recall, the phase $\varphi$ varies as $e^{-i\varphi N}$.

	In addition to this sufficient entanglement test, we can directly determine the entanglement quasiprobabilities for the dephased N00N state following the same procedure as formulated in Sec. \ref{ch:pureN00N}.
	The analytical solution in the Appendix yields the following quasiprobability:
	\begin{align}\label{eq:EQDelta}
		P_{\rm Ent}&\cong
		\frac{1}{4}(0,0,2,2,\lambda,\lambda,\lambda,\lambda,-\lambda,-\lambda,-\lambda,-\lambda).
	\end{align}
	Compared to Eq. \eqref{eq:QuasiProbPure} for the case of a pure N00N state, we observe that the last eight entries are diminished by $\lambda\leq 1$.
	Those entries relate to similar solutions to those in Eq. \eqref{eq:nontrivial}, which address the interferences.
	However, we also immediately observe that $P_{\rm Ent}$ always contains negative entries, $-\lambda<0$.

	The behavior of these entries in dependence on the Gaussian parameter $\delta$ is shown as the solid curve in Fig. \ref{fig:depha}.
	Although these negativities decay with increasing $\delta$, they only vanish in the limit of a full dephasing, i.e., $\delta\to\infty$.
	Therefore, we can conclude that for any finite dephasing the entanglement of the N00N state is preserved, as the negativities of $P_{\rm Ent}$ are necessary and sufficient for entanglement.
	Compared to the witness-related approach (dashed curve in Fig. \ref{fig:depha}), we can see that the entanglement quasiprobabilities detect entanglement in a wider range
	than the witnesses test.
	This is not surprising as the method of entanglement quasiprobabilities is a full state representation and uncovers all the entanglement, while the witness approach depends on the choice of the test operator, which might not be optimal.

\subsection{Fluctuating, atmospheric losses}

	The second imperfection to be studied is the case of N00N states suffering from fluctuating losses.
	As in the case of dephasing, we consider both, entanglement witness and quasiprobabilities, in order to characterize the entanglement.
	Furthermore, we illustrate how choosing a different test operator can enhance the region of detected entanglement.

	For simplicity, we restrict ourselves to $N=2$ and correlated fluctuating losses, which yields the state
	\begin{align}\label{eq:fluctuatingN00N}
	\begin{aligned}
		\hat\rho_{\rm fluc}=&
		\langle(1- T^2)^2\rangle|0,0\rangle\langle 0,0|
		\\&+\langle T^2(1-T^2)\rangle(|1,0\rangle\langle 1,0| +|0,1\rangle\langle 0,1|)
		\\&+\langle T^4\rangle|\psi_2\rangle\langle\psi_2|
	\end{aligned}
	\end{align}
	[see  Eq. \eqref{eq:PDTCchannel}].
	Correlated fluctuating losses means that both optical modes copropagate through an atmospheric channel, which results in identical randomizations of the transmission coefficients, $T_a=T_b=T$ (see \cite{BSSV2016,BSSV2017}).
	Note that the state is then determined by the two transmission moments $\langle T^2\rangle$ and $\langle T^4\rangle$ and it is a mixture of separable states and the N00N state $|\psi_2\rangle$.

\begin{figure}[hb]
	\includegraphics[width=.95\columnwidth]{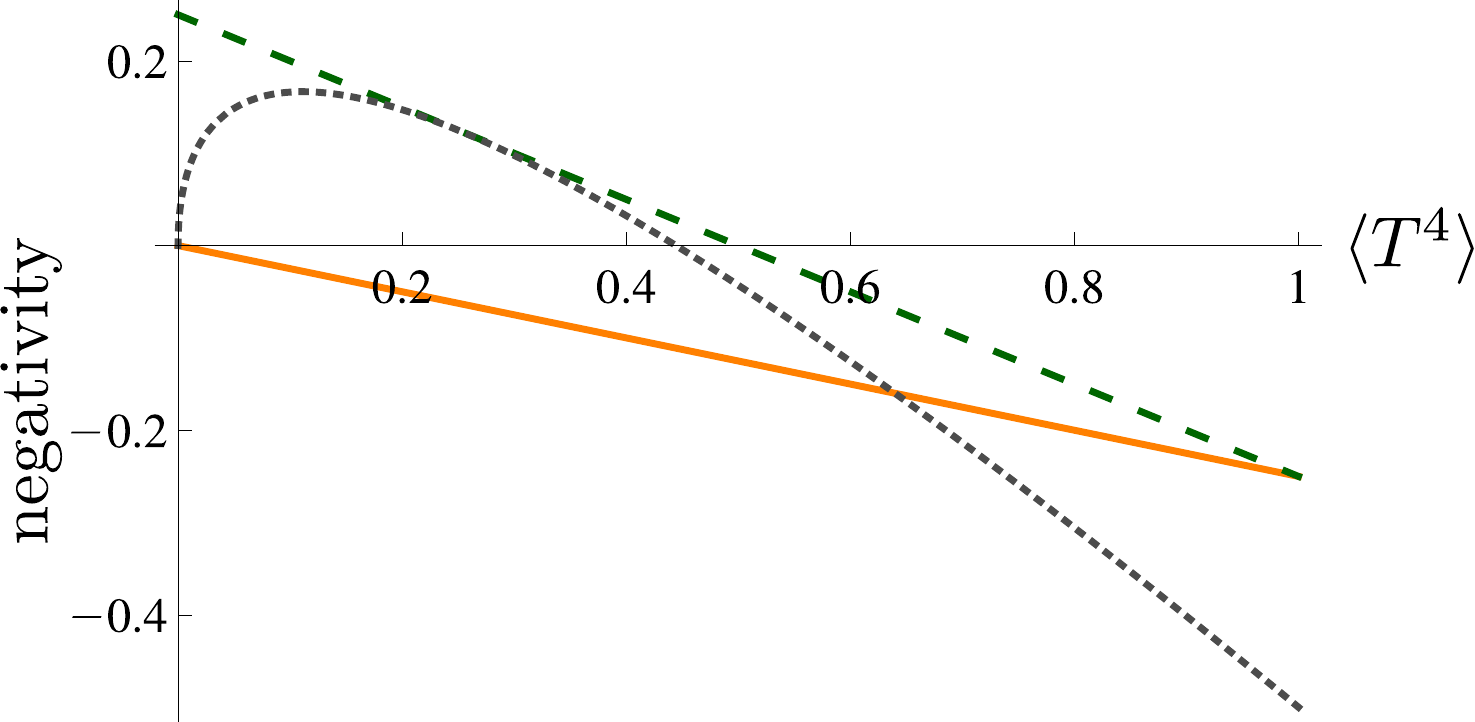}
	\caption{(Color online)
		Entanglement of a N00N state under correlated fluctuating loss, characterized via the fourth-order moment $\langle T^4\rangle$ of the probability distribution of the transmission coefficient $T$.
		Negative values certify entanglement. 
		The dashed and dotted lines correspond to the witness conditions in Eqs. \eqref{eq:WitnessLoss} and \eqref{eq:WAtilde}, respectively.
		The solid line shows the negativity of the quasiprobability in Eq. \eqref{eq:PentLoss}.
	}\label{fig:loss}
\end{figure}

	In order to access the entanglement of the state \eqref{eq:fluctuatingN00N}, we start again with the witness condition in Eq. \eqref{eq:EC}, resulting in the entanglement condition
	\begin{align}\label{eq:WitnessLoss}
		\frac{1}{4}-\frac{\langle T^4\rangle}{2}<0.
	\end{align}
	This implies entanglement if $\langle T^4\rangle>1/2$, which limits the maximal amount of sustainable loss.
	The test \eqref{eq:WitnessLoss} is plotted in Fig. \ref{fig:loss} (dashed line) in dependence on the loss parameter $\langle T^4\rangle$.

	To demonstrate the influence of the choice of the test operator $\hat L$ on the detectable range of entanglement, let us now consider the modified test operator,
	\begin{align}\label{eq:Atilde}
		\hat{L}=|N,0\rangle\langle 0,N|+|0,N\rangle\langle N,0|+\frac{1}{2}|0,0\rangle\langle 0,0|,
	\end{align}
	for which we can formulate the corresponding witness condition.
	Again, from the solutions in the Appendix follows that the maximal separability eigenvalue is $\sup_i\{g_i\}=1/2$.
	This allows us to formulate the following entanglement condition:
	\begin{align}\label{eq:WAtilde}
		\frac{1}{2}-\langle\hat L\rangle<0.
	\end{align}
	For the state in Eq. \eqref{eq:fluctuatingN00N}, this explicitly reads
	\begin{align}\label{eq:WAtildeN00N}
		\langle T^2\rangle-\frac{3}{2}\langle T^4\rangle<0.
	\end{align}

	For the deterministic loss case, $\langle T^4\rangle=T^4=\langle T^2\rangle^2$, we certify entanglement for $T^2>2/3$.
	The condition in Eq. \eqref{eq:WAtildeN00N} is shown in Fig. \ref{fig:loss} (dotted line) for such a deterministic loss.
	The range of a successful entanglement verification covers a wider range of parameters than provided by the previous test \eqref{eq:WitnessLoss} (dotted line).
	This enhancement originates from the larger overlap of the witness with the state understudy, i.e., $\langle \hat L\rangle$ increases, while the  maximal separability eigenvalue is not affected and remains constant.
	Moreover, the scenario of turbulent loss, $\langle T^4\rangle\neq \langle T^2\rangle^2$, together with the entanglement condition \eqref{eq:WAtildeN00N} gives the bound $\langle T^4\rangle/\langle T^2\rangle>2/3.$

	The witness approach allows us to characterize fluctuating loss conditions for which entanglement can be preserved.
	Similar to the considered examples, this also applies to the general case of noisy N00N states.
	Such an analysis is of great importance for free-space quantum communication.
	Firstly, it renders it possible to determine which channel losses are acceptable in order to preserve entanglement.
	Secondly, one is able to construct optimized entanglement conditions for various fluctuating loss conditions.

	Finally, let us calculate the entanglement quasiprobabilities for the N00N state undergoing correlated fluctuating losses.
	Similarly to the previously considered cases, we can exactly derive the entanglement quasiprobabilities for all cases of fluctuating losses,
\begin{widetext}
	\begin{align}\label{eq:PentLoss}
		&P_{\rm Ent}\cong\\
		&\left(
			\langle(1-T^2)^2\rangle,0,
			\frac{\langle T^4\rangle}{2},\frac{\langle T^4\rangle}{2},
			-\frac{\langle T^4\rangle}{4},-\frac{\langle T^4\rangle}{4},-\frac{\langle T^4\rangle}{4},-\frac{\langle T^4\rangle}{4},\frac{\langle T^4\rangle}{4},
			\frac{\langle T^4\rangle}{4},\frac{\langle T^4\rangle}{4},\frac{\langle T^4\rangle}{4},
			\langle T^2(1-T^2)\rangle,\langle T^2(1-T^2)\rangle
		\right).\nonumber
	\end{align}
\end{widetext}
	The entries in $P_{\rm Ent}$ are associated with the vectors in Eqs. \eqref{eq:trivial} and \eqref{eq:nontrivial}, including the new, last two entries which resemble the separable contributions $|0,1\rangle$ and $|1,0\rangle$.
	As $P_{\rm Ent}$ exhibits the negative entries $-\langle T^4\rangle/4$, entanglement can be verified for any $\langle T^4\rangle\neq0$.
	The behavior of these negativities in dependence on $\langle T^4\rangle$ are shown as the solid line in Fig. \ref{fig:loss}.
	Hence, based on the quasiprobability distribution \eqref{eq:PentLoss}, we can conclude that any correlated (turbulent) loss channel with a nonzero transmission preserves the entanglement of the N00N state.
	A similar loss-robust behavior was indeed observed in other scenarios, e.g., for entangled states in Mach-Zehnder interferometers \cite{Gholipour2016} or continuous-variable quantum states \cite{Buono2012}, and it is also predicted for the N00N state here.

\section{Multipartite generalization and comparison with nonclassical correlations}\label{ch:Generalization}

	To get a deeper understanding of our findings, let us generalize and compare our previous results.
	First, we consider a generalization of the witnessing approach to multipartite noisy N00N states for which we give an example.
	In this context, we discuss the limitations of the partial-transposition approach compared to our method.
	Second, we compare our entanglement quasiprobabilities with the bipartite Glauber-Sudarshan quasiprobability distribution.

\subsection{Multimode entanglement}

	A straightforward $d$-partite generalization of the N00N state is defined as
	\begin{equation}\label{eq:MultiNo00N}
		|\psi_{N,d}\rangle=\frac{1}{\sqrt{d}}\sum\limits_{m=1}^d |0\rangle_1|0\rangle_2\cdots|N\rangle_m\cdots|0\rangle_d,
	\end{equation}
	with $d\geq 2$ modes (see, e.g., Ref. \cite{Humphreys2013}).
	The state $|\psi_{N,d}\rangle$ is also known as the generalized $W$ state, where the ordinary tripartite $W$ state reads $(|1\rangle_1|0\rangle_2|0\rangle_3+|0\rangle_1|1\rangle_2|0\rangle_3+|0\rangle_1|0\rangle_2|1\rangle_3)/\sqrt3$.
	Also, $d=2$ yields the previously studied bipartite N00N state.
	We are now interested in the entanglement properties of such multimode N00N states, while restricting to $d=3$ modes.
	In the multimode scenario, we can have different forms of separable states, leading to different notions of inseparability (i.e., entanglement).
	Fully separable states are based on products of the form $|a\rangle_1|b\rangle_2|c\rangle_3$.
	By contrast, partially separable states require the separation of one mode, for example, $|a\rangle_1|\phi\rangle_{2,3}$, where $|\phi\rangle_{2,3}$ could be an entangled state in the last two modes.
	Because of the symmetry of the state \eqref{eq:MultiNo00N}, it does not matter which mode is separated.

\begin{figure}[hb]
	\includegraphics[width=.95\columnwidth]{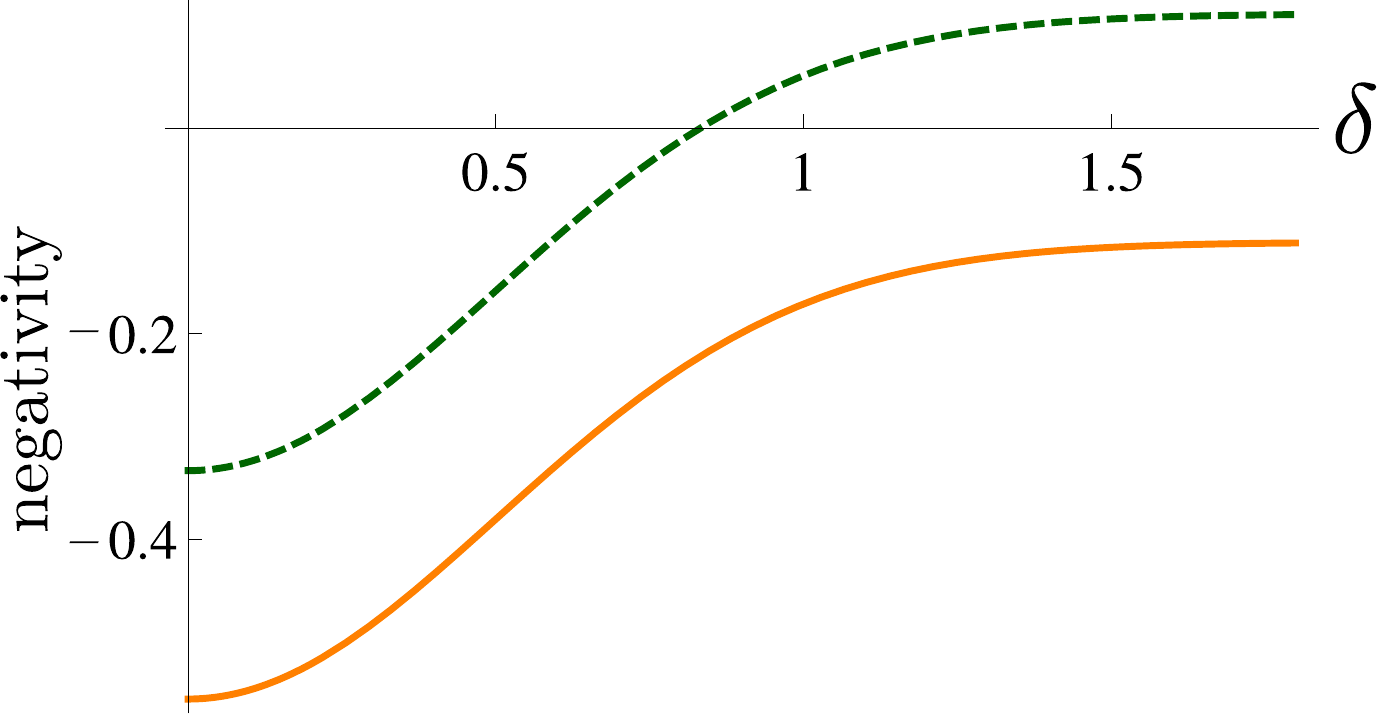}
	\caption{(Color online)
		The expectation value of the witness \eqref{eq:Wthreemode} for partial (solid line) and full (dashed line) separability of a tripartite N00N state with dephasing of strength $\delta$ in one mode.
		A negative value certifies the presence of the corresponding form of inseparability.
	}\label{fig:mdephasing}
\end{figure}

	In order to identify multimode entanglement, one can employ the multipartite SEP \cite{Sperling2013}.
	For example, this gives the three-mode entanglement witness
	\begin{equation}\label{eq:Wthreemode}
		\hat W_{\rm full/part}=f_{\rm full/part}-|\psi_{N,3}\rangle\langle \psi_{N,3}|.
	\end{equation}
	Similarly to our demonstration for the bipartite case, we get the bounds $f_{\rm full}=2/3$ and $f_{\rm part}=4/9$ \cite{Sperling2013,Pagel2013}.
	A state is not fully separable if $\langle \hat W_{\rm full}\rangle<0$ and not partially separable if $\langle \hat W_{\rm part}\rangle<0$.
	For instance, we can consider the state $|\psi_{N,d}\rangle\langle \psi_{N,d}|$ which undergoes Gaussian dephasing in one of the modes (see also Sec. \ref{ch:Dephasing}).
	This yields $\langle\hat W_{\rm part}\rangle=-(4\lambda+1)/9$ and $\langle\hat W_{\rm full}\rangle=-(4\lambda-1)/9$, where $\lambda=\exp(-\delta N^2/2)$.
	The result of the two conditions is shown in Fig. \ref{fig:mdephasing}.
	Surprisingly, the dephased three-mode N00N state always violates the condition for partial separability (solid line), which is not true for the case of full separability (dashed line).

	Obviously, the situation becomes ever more complex for increasing numbers of modes $d$, because we can consider separations beyond bi- and triseparable states.
	The experimental implementation of our approach to construct entanglement tests has successfully verified such different, complex forms of multipartite entanglement \cite{Gerke2015,Gerke2016}.
	By contrast, the popular positive partial transposition criterion \cite{Perez1996} cannot perform such a task.
	Since any mode is either transposed or not, this effectively decomposes the system into bipartitions and, thus, can only identify entanglement based on  bipartitions. 
	Using our technique, we could additionally verify full inseparability for $d=3$ modes, which defines tripartite entanglement, and thus outperforms the partial transposition.

\subsection{Nonclassical correlations}

	Furthermore, let us discuss the conceptual importance of the entanglement quasiprobabilities representation \eqref{eq:Pent} and its relation to other notions of nonclassical correlations.
	Entanglement quasiprobabilities provide a direct connection to the very definition of separable states.
	By definition, any bipartite separable state can be given by a convex combination of product states \cite{Werner89},
	\begin{align}\label{eq:separable}
		\hat \sigma=\sum_k p_k|a_k,b_k\rangle\langle a_k,b_k|,
	\end{align}
	with $p_k$ being a classical probability distribution.
	Any bipartite state which cannot be expressed in this form is entangled.
	However, entangled states can be decomposed similarly to Eq. \eqref{eq:separable} if we allow for negativities in the probability distribution [cf. Eq. \eqref{eq:Pent}].
	Therefore, we immediately observe the structural similarity between the definition of separability and inseparability, and how separability is overcome in terms of quasiprobabilities for entangled states. This demonstrates the fundamental character of such a quasiprobability representation.

	Moreover, we would also like to emphasize the conceptual similarity to the Glauber-Sudarshan $P$ representation \cite{Glauber,Sudarshan}.
	The two-mode $P$ representation is the expansion of the considered quantum state in a diagonal coherent-state basis
	\begin{align}\label{eq:Pfunction}
		\hat \rho=\int d^{2}\alpha d^{2}\beta P(\alpha,\beta)|\alpha,\beta\rangle\langle \alpha,\beta|.
	\end{align}
	The standard definition of a classical state in quantum optics is that $P$ is a valid probability distribution, while nonclassical quantum states do not allow for such a non-negative $P$ distribution \cite{Titulaer1965}.
	Therefore, the Glauber-Sudarshan $P$ representation and the entanglement quasiprobabilities have a similar structure and can be interpreted in the same way.
	When comparing this to our entanglement quasiprobability representation \eqref{eq:Pent}, we find similarities and differences.
	In both cases, $P_\mathrm{ent}(a,b)\geq0$ and $P(\alpha,\beta)\geq0$ identify classical properties with respect to different classical references.
	However, the $P$ representation in quantum optics relies on the preferred basis of coherent states $|\alpha,\beta\rangle$.
	The entanglement quasiprobabilities are independent of the local representation, since any product state $|a,b\rangle$ can be used [see Eq. \eqref{eq:separable}].
	The other way around, a classical Glauber-Sudarshan representation also implies separability \cite{Sperling2009QP}.

	Let us consider the case of a fully dephased N00N state [see Eq. \eqref{eq:dephasing} for $\lambda=0$].
	This state can be given by its $P$ representation
	\begin{align}
	\begin{aligned}
		P(\alpha,\beta)&=\sum\limits_{k=0}^N \binom{N}{k}\frac{1}{2(N-k)!}\partial_\alpha^{N-k}\partial_{\alpha^*}^{N-k}\delta(\alpha)\delta(\beta)\\
		&+\sum\limits_{k=0}^N \binom{N}{k}\frac{1}{2(N-k)!}\partial_\beta^{N-k}\partial_{\beta^*}^{N-k}\delta(\alpha)\delta(\beta),
	\end{aligned}
	\end{align}
	where $\delta$ is the Dirac-delta distribution.
	As $P(\alpha,\beta)$ exhibits finite orders of derivatives of the $\delta$ distribution \cite{Sperling2016a}, $P(\alpha,\beta)$ is not positive semidefinite.
	Therefore, the state $(|N,0\rangle\langle N,0|+|0,N\rangle\langle 0,N|)/2$ is quantum correlated in terms of the Glauber-Sudarshan $P$ function.
	Still, the entanglement quasiprobability representation \eqref{eq:EQDelta} of this separable state is naturally non-negative, demonstrating the absence of quantum correlations in terms of entanglement.
	Hence, our approach of entanglement quasiprobabilities allows for discerning entanglement from the notions of nonclassicality in quantum optics, while being based on a comparable approach.
	It is also worth mentioning that the general basis independence of entanglement has been also discussed in Ref. \cite{Eliza2017}, and another quasiprobability representation has been introduced in Ref. \cite{Ryu2013}.

\section{Conclusions}\label{ch:Summary}

	In summary, we presented a fully analytical study which allows one to characterize the entanglement properties of noisy N00N states.
	We used two methods to determine entanglement, namely entanglement witnesses and entanglement quasiprobabilities.
	Both methods are based on the solutions of the separability eigenvalue equations.
	We solved those equations for the considered family of states.
	As examples, we investigated the impact of dephasing and turbulent losses on the entanglement of N00N states.
	Still, our general solution directly applies to any noisy N00N state, which includes much more general scenarios of imperfections.
	
	We compared the entanglement quasiprobabilities with the witness approach.
	By construction, the entanglement quasiprobabilities are capable of uncovering any kind of entanglement of the states under study.
	Therefore, the entanglement quasiprobabilities represent a useful tool for the necessary and sufficient entanglement verification.
	However, to date, it is unclear how to apply this technique experimentally.
	In contrast, entanglement witnesses have been successfully applied to experimental data in the past.
	Hence, they present a sufficient entanglement condition only.
	Independently of the preferred case (either necessary and sufficient or experimentally accessible conditions), our method can be applied.

	We applied both approaches and, specifically, characterized N00N states that are subjected to dephasing and fluctuating, atmospheric losses.
	We started our entanglement analysis with pure N00N states and included both types of imperfections.
	We could demonstrate via the entanglement quasiprobabilities that the entanglement of the initially pure N00N states can be detected for any dephasing and loss---excluding the singular cases of a full phase diffusion and a zero transmittance.
	Furthermore, we investigated for which imperfections the entanglement can still be detected with measurable witnesses.

	Moreover, we studied the multipartite generalization and showed that our witnessing approach exceeds the capabilities of the partial transposition.
	In particular, we can clearly distinguish the cases of full and partial separability, which is important for classifying the entanglement among multipartite quantum networks and channels.
	By relating the entanglement quasiprobabilities to the Glauber-Sudarshan representation, we have also been able to discuss the role of different forms of quantum correlations.
	This allows for distinguishing the general quantum phenomena, defined through the two-mode phase-space representation, into one part that covers separable forms of nonclassicality and one part with inseparable quantum effects.

	In conclusion, we implemented two complementary approaches to access the entanglement of noisy N00N state via entanglement witnesses and entanglement quasiprobabilities.
	They enabled us to systematically and analytically infer how many imperfections can be tolerated in general or for certain measurements.
	This renders it possible to predict stability requirements of dephasing channels or favorable turbulence conditions of atmospheric communication links.

\begin{acknowledgments}
	The authors gratefully acknowledge financial support by Deutsche Forschungsgemeinschaft through Grant No. VO 501/22-1.
\end{acknowledgments}

\appendix*

\section{Exact solutions}\label{app:SEPsolution}

	In this Appendix, we provide a detailed calculation which yields the solution of separability eigenvalue equations \eqref{eq:SEP} \cite{Sperling2009,Sperling2013}.
	We consider the test operators \eqref{eq:testop}, which have the same structure as the noisy N00N states \eqref{eq:NoisyN00N}, in an infinite-dimensional, bipartite Hilbert space.
	First solutions for $|a\rangle$ and $|b\rangle$ can be directly found by making the ansatz of states being either parallel to $|0\rangle$ or perpendicular to $|0\rangle$ (cf. first four rows in Table \ref{tab:trivSol}).

\begin{table}[ht]
	\caption{
		Solutions $(g,|a,b\rangle)$ of the SEP in Eq. \eqref{eq:SEP} for the operators in Eq. \eqref{eq:testop}.
		$i$ is a positive integer; $|x\rangle$ and $|y\rangle$ are arbitrary vectors perpendicular to $|0\rangle$ (e.g., the Fock states $|x\rangle=|y\rangle=|i\rangle$).
		It holds that $\vartheta=\varphi-\arg\gamma_i$ (e.g., one can choose multiples of a right angle, $\varphi=0,\pi/2,\pi,3\pi/2$); $\mu$ and $\nu$ are explicitly given in Eq. \eqref{eq:munuFin}.
	}\label{tab:trivSol}
	\begin{tabular}{cc}
		\hline\hline
		$g$ & $|a,b\rangle$ \\
		\hline
		$L^{(0)}$ & $|0\rangle{\otimes}|0\rangle$ \\
		$L^{(A)}_i$ & $|i\rangle{\otimes}|0\rangle$ \\
		$L^{(B)}_i$ & $|0\rangle{\otimes}|i\rangle$ \\
		$0$ & $|x\rangle{\otimes}|y\rangle$ \\
		$
			\frac{\mu\nu}{\mu{+}\nu{-}L^{(0)}}
		$ & $
			\frac{\sqrt\nu|0\rangle{+}\sqrt{\mu{-}L^{(0)}}e^{\iota\varphi}|i\rangle}{\sqrt{\nu{+}\mu{-}L^{(0)}}}
			{\otimes}
			\frac{\sqrt\mu|0\rangle+\sqrt{\nu{-}L^{(0)}}e^{\iota\vartheta}|i\rangle}{\sqrt{\nu{+}\mu{-}L^{(0)}}}
		$\\
		\hline\hline
	\end{tabular}
\end{table}

	In the case of superpositions of parallel and perpendicular parts, we can use a parametrization
	\begin{align}\label{eq:superposSol}
		|a\rangle=\frac{|0\rangle+|x\rangle}{\sqrt{1+\langle x|x\rangle}}
		\quad
		\text{and}
		\quad
		|b\rangle=\frac{|0\rangle+|y\rangle}{\sqrt{1+\langle y|y\rangle}},
	\end{align}
	where $|x\rangle,|y\rangle\perp|0\rangle$ and $|x\rangle,|y\rangle\neq 0$.
	Inserting decomposition \eqref{eq:superposSol} into the first and second Eqs. \eqref{eq:SEP}, separating for contributions parallel and perpendicular to $|0\rangle$, and using a proper scaling, we get
	\begin{subequations}
	\begin{align}\label{eq:SEPinsert_a}
		L^{(0)}+\langle y|\left(\hat L^{(B)}|y\rangle+\hat\gamma^\dag|x\rangle\right)=&g\left(1+\langle y|y\rangle\right),
		\\\label{eq:SEPinsert_b}
		\hat L^{(A)}|x\rangle+\hat\gamma|y\rangle=&g\left(1+\langle y|y\rangle\right)|x\rangle,
		\\\label{eq:SEPinsert_c}
		L^{(0)}+\langle x|\left(\hat L^{(A)}|x\rangle+\hat\gamma|y\rangle\right)=&g\left(1+\langle x|x\rangle\right),
		\\\label{eq:SEPinsert_d}
		\hat L^{(B)}|y\rangle+\hat\gamma^\dag|x\rangle=&g\left(1+\langle x|x\rangle\right)|y\rangle,
	\end{align}
	\end{subequations}
	where we defined the single-mode operators ($S=A,B$)
	\begin{align}\label{eq:1ModeOp}
		\hat L^{(S)}=\sum_{i=1}^\infty L^{(S)}_i|i\rangle\langle i|
		\text{ and }
		\hat \gamma=\sum_{i=1}^\infty \gamma_i|i\rangle\langle i|.
	\end{align}
	Note that $\hat L^{(A)}$, $\hat L^{(B)}$, and $\hat \gamma$ commute and that they have the null space (i.e., kernel) spanned by $|0\rangle$.

	Inserting Eqs. \eqref{eq:SEPinsert_d} and \eqref{eq:SEPinsert_b} into \eqref{eq:SEPinsert_a} and \eqref{eq:SEPinsert_c}, respectively, we obtain two identical equations,
	\begin{align}\label{eq:gUni}
		L^{(0)}=g(1-\langle x|x\rangle\langle y|y\rangle).
	\end{align}
	In the continuation of our calculation, it is convenient to introduce the real-valued parameters
	\begin{align}\label{eq:munu}
		\mu=g(1+\langle x|x\rangle)
		\quad
		\text{and}
		\quad
		\nu=g(1+\langle y|y\rangle).
	\end{align}
	Resolving those equations for $\langle x|x\rangle$ and $\langle y|y\rangle$ and inserting the result into Eq. \eqref{eq:gUni} yields the separability eigenvalue in the form
	\begin{align}
		g=\frac{\mu\nu}{\mu+\nu-L^{(0)}}.
	\end{align}
	Using this representation and Eq. \eqref{eq:munu}, we can write
	\begin{align}\label{eq:AmplitudesPre}
		\langle x|x\rangle=\frac{\mu-L^{(0)}}{\nu}
		\quad
		\text{and}
		\quad
		\langle y|y\rangle=\frac{\nu-L^{(0)}}{\mu}.
	\end{align}

	In the next step, Eqs. \eqref{eq:SEPinsert_b} and \eqref{eq:SEPinsert_d} can be combined into a block form
	\begin{align}\label{eq:MatrixForm}
		\begin{pmatrix}
			0 \\ 0
		\end{pmatrix}
		=
		\begin{pmatrix}
			\hat L^{(A)}-\nu & \hat \gamma \\ \hat \gamma^\dag & \hat L^{(B)}-\mu
		\end{pmatrix}
		\begin{pmatrix}
			|x\rangle \\ |y\rangle
		\end{pmatrix}.
	\end{align}
	To solve this equation, one has to compute the null space of the block-operator which requires a vanishing determinant.
	For the operators defined in Eq. \eqref{eq:1ModeOp}, this results in the constraint
	\begin{align}\label{eq:characteristic}
		0=\prod_{i=1}^\infty \left([L_i^{(A)}-\nu][L_i^{(B)}-\mu]-|\gamma_i|^2\right),
	\end{align}
	which factorizes into terms for each $i>0$.
	Therefore, we can solve the problem for each $i$ individually which further implies that
	\begin{align}\label{eq:Fock}
		|x\rangle=\alpha|i\rangle
		\quad
		\text{and}
		\quad
		|y\rangle=\beta|i\rangle,
	\end{align}
	with $\alpha,\beta\in\mathbb C$ and the Fock state $|i\rangle$.
	Inserting $|x\rangle$ and $|y\rangle$ into Eq. \eqref{eq:MatrixForm}, we get the following relation between $\alpha$ and $\beta$:
	\begin{align}\label{eq:relateAlphaBeta}
		\frac{\beta}{\alpha}=\frac{\gamma_i^\ast}{\mu-L_i^{(B)}}=\frac{\nu-L_i^{(A)}}{\gamma_i}.
	\end{align}
	Thus, we get for the phases $\arg\beta=\arg\alpha-\arg\gamma_i.$
	\begin{subequations}
	For the amplitudes, we get from Eq. \eqref{eq:relateAlphaBeta}
	\begin{align}\label{eq:ratio1}
		\frac{\beta}{\alpha}\frac{\beta^\ast}{\alpha^\ast}=\frac{\nu-L_i^{(A)}}{\mu-L_i^{(B)}}.
	\end{align}
	The combination of Eqs. \eqref{eq:AmplitudesPre} and \eqref{eq:Fock} yields
	\begin{align}\label{eq:ratio2}
		\frac{|\beta|^2}{|\alpha|^2}=\frac{\nu(\nu-L^{(0)})}{\mu(\mu-L^{(0)})}.
	\end{align}
	\end{subequations}

	Equating Eq. \eqref{eq:ratio1} with Eq. \eqref{eq:ratio2} and using the condition that the $i$th factor in Eq. \eqref{eq:characteristic} is zero, we find the following system of equations:
	\begin{align}
	\begin{aligned}
		&\nu\frac{\nu-L^{(0)}}{\nu-L_i^{(A)}}=\mu\frac{\mu-L^{(0)}}{\mu-L_i^{(B)}},\\
		&[\nu-L_i^{(A)}][\mu-L_i^{(B)}]=|\gamma_i|^2.
	\end{aligned}
	\end{align}
	This problem can be straightforwardly solved:
	\begin{align}\label{eq:munuFin}
	\begin{aligned}
		\mu=&\frac{L^{(A)}_iq^{(B,A)}\pm|\gamma_i|\sqrt{q^{(A,B)}q^{(B,A)}}}{q^{(A,A)}},\\
		\nu=&\frac{L^{(B)}_iq^{(A,B)}\pm|\gamma_i|\sqrt{q^{(A,B)}q^{(B,A)}}}{q^{(B,B)}},
	\end{aligned}
	\end{align}
	where we apply the following definition ($S,T\in\{A,B\}$):
	\begin{align}
	\begin{aligned}
		q^{(S,T)}=&L_i^{(S)}(L_i^{(T)}-L^{(0)})-|\gamma_i|^2.
	\end{aligned}
	\end{align}
	From Eq. \eqref{eq:AmplitudesPre}, we also get
	\begin{align}
		|\alpha|^2=\frac{\mu-L^{(0)}}{\nu}
		\quad
		\text{and}
		\quad
		|\beta|^2=\frac{\nu-L^{(0)}}{\mu}.
	\end{align}
	In conclusion, the separability eigenvector $|a,b\rangle$ is composed of the states in Eq. \eqref{eq:superposSol}, which can also be written in the form
	\begin{align}
	\begin{aligned}
		|a\rangle&=\frac{
			\sqrt\nu|0\rangle+\sqrt{\mu-L^{(0)}}e^{\iota\varphi}|i\rangle
		}{\sqrt{
			\nu+\mu-L^{(0)}
		}},
		\\
		|b\rangle&=\frac{
			\sqrt\mu|0\rangle+\sqrt{\nu-L^{(0)}}e^{\iota\vartheta}|i\rangle
		}{\sqrt{
			\mu+\nu-L^{(0)}
		}}.
	\end{aligned}
	\end{align}
	Here, $\iota$ denotes the imaginary unit to avoid confusions with the positive integer $i$.
	The phases $\varphi$ and $\vartheta$ have to be chosen such that Eq. \eqref{eq:relateAlphaBeta} is satisfied ($\vartheta=\varphi-\arg\gamma_i$).
	The separability eigenvalue is expressed in terms of $\mu$ and $\nu$ in Eq. \eqref{eq:gUni}.
	Therefore, the SEP is solved for the test and density operators under study.
	All solutions together with the trivial cases can be found in Table \ref{tab:trivSol}.


\end{document}